\documentstyle[aps,prb,twocolumn]{revtex}
\begin{document}
\draft
\title{Many-Spin Interactions and Spin Excitations in Mn$_{12}$}
\author{M.\ I.\ Katsnelson}
\address{Institute of Metal Physics, Ekaterinburg 620219, Russia}
\author{V.\ V.\ Dobrovitski and B.\ N.\ Harmon}
\address{Ames Laboratory, Iowa State University, Ames, Iowa 50011}
\date{\today}
\maketitle
\begin{abstract}
In this work, the many-spin interactions taking place in Mn$_{12}$ 
large-spin clusters are extensively studied using the 8-spin model 
Hamiltonian, for which we determine the possible parameters based 
on experimental data.
Account of 
the many-spin excitations satisfactorily  explains 
positions of 
the neutron scattering peaks, results of EPR measurements and 
the temperature dependence of magnetic susceptibility. 
In particular, strong Dzyaloshinsky-Morya interactions are 
found to be important for description of neutron scattering data.
The role of these interactions for the relaxation of the 
magnetization is qualitatively discussed. 
\end{abstract}


\section*{Introduction}

In the last years, a new kind of magnetic compounds, the 
magnetic molecules, has been drawing the 
attention of 
physicists as well as chemists \cite{genintro}. 
Such molecules each contain a 
large number (typically, 10 to 20) of paramagnetic 
ions (such as Mn, Fe or Cu) coupled by exchange interactions. 
Each molecule,
therefore, presents a mesoscopic system that is neither 
totally microscopic, nor totally macroscopic, but where 
micro- and macroscopic behavior coexist. 
These materials are promising for various practical 
applications \cite{appl}. On the other hand, the coexistence 
of quantum and classical behavior in the clusters makes 
them very suitable objects for study of 
macroscopic quantum effects in spin systems \cite{leggett,conf}.
These studies, clarifying many problems of quantum 
theory of measurements \cite{menskii}, are also important 
for development of a physical basis for practical 
implementation of powerful algorithms of quantum computations,
quantum cryptography and quantum searching \cite{quanta}.

Particularly, the 
$\rm Mn_{12}O_{12}(CH_3COO)_{16}(H_2O)_4$ mo\-le\-cu\-les 
(below referred to as Mn$_{12}$)
recently became a subject of great interest. 
Each molecule \cite{gatteschi,gat1}
contains a cluster of twelve manganese ions 
surrounded by acetate radicals and water molecules.
The ground state of the clusters corresponds to a 
large total spin ${\cal S}=10$.
The clusters possess a strong easy-axis anisotropy: 
the zero-field 
splitting between the states with ${\cal S}_z=\pm 10$ 
and ${\cal S}_z=\pm 9$ (where ${\cal S}_z$ is the 
value of $z$-projection of the total cluster spin)
is 14.4 K. 
Being stacked into a crystal, the molecules form 
a tetragonal lattice; in so doing the magnetic interactions 
between different clusters are very small (of order of 
10$^{-2}$ T).
Thus, the crystal consisting of these molecules 
can be considered as an assembly of ideal noninteracting 
superparamagnetic entities, each being identical to 
the others.

These clusters have been 
successfully used for the study of mesoscopic quantum effects.
In particular, resonant magnetization tunneling 
has been unambigously registered in 
experiments on Mn$_{12}$ \cite{jumps,jumps1}. 
Moreover, there are 
experimental results \cite{barrel,park} 
supporting the hypothesis of 
"ground state-to-ground state" tunneling in Mn$_{12}$
below 2 K (for more detailed discussion, see 
Sec.\ \ref{sectrelax}).

However, the progress in understanding the physical 
properties of Mn$_{12}$ 
is greatly hampered by the lack of an adequate description 
of these clusters. Indeed, the description of Mn$_{12}$ 
as a single spin ${\cal S}=10$ entity has been the starting point in
most works devoted to this subject. We know of only a few 
theoretical attempts to account for the internal spin structure 
of the cluster \cite{gat1,gat2,zvezdin}, but even in these the 
relativistic anisotropic interactions 
have not been taken into account.
In view of recent experiments \cite{neutr,specheat,dmbar} 
showing that the 
single-spin model is seriously deficient, it is worthwhile 
reconsidering the many-spin aspects of Mn$_{12}$.

In this paper we focus on the 
many-spin interactions in Mn$_{12}$ clusters. 
We account for not only isotropic exchange 
interactions between ions in the cluster, but also 
various anisotropic interactions possibly present in Mn$_{12}$.
Based on the results, we propose a spin Hamiltonian
for these clusters. We show that this Hamiltonian can 
reproduce satisfactorily most recent experimental 
results, such as 
positions of neutron scattering peaks, high-frequency EPR 
data and the experimental dependence of the magnetic susceptibility 
on temperature. We note that the account of anisotropic 
interactions, especially the Dzyaloshinsky-Morya interaction
(which has been missing up to now), is crucial for 
a detailed description of the experimental data.

The paper is organized as follows. In Sec.\ \ref{sectmod} 
we describe the basic model of Mn$_{12}$ used in this work 
and establish roughly its domain of validity.
In Sec.\ \ref{sectham} we derive and discuss the spin 
Hamiltonian for this model.
Sec.\ \ref{sectexpres} is devoted 
to discussion of relevant experimental data. In 
Sec.\ \ref{sectparam} the numerical procedure 
used for calculations is discussed and the possible 
parameters of the spin Hamiltonian are presented. 
Comparison with experimental data is made.
The results obtained are analyzed qualitatively and 
discussed in Sec.\ \ref{sectinter}, where the interpretation 
of the neutron scattering data is presented.
In Sec.\ \ref{sectrelax} we qualitatively 
discuss the relation between Dzyaloshinsky-Morya interactions 
and magnetic relaxation in Mn$_{12}$. A summary is provided 
in Sec.\ \ref{sectsum}.

\section{The dimerized 8-spin model of M\lowercase{n}$_{12}$}
\label{sectmod}

The cluster Mn$_{12}$, schematically shown in Fig.\ 
\ref{figmn1}, consists of eight Mn$^{3+}$ ions 
having the spin 2 and four Mn$^{4+}$ ions having the 
spin 3/2. The ions are coupled by exchange interactions, 
indicated in Fig.\ \ref{figmn1} by different lines connecting 
the ions. The values of the exchange integrals are not known, but 
estimates are given in Ref.\ \onlinecite{gat1}: 
$J_1=-150$ cm$^{-1}$ (AFM exchange),
$J_2=J_3=-60$ cm$^{-1}$ and 
$|J_4|<30$ cm$^{-1}$. These values are rough, but 
describe correctly the scale of exchange 
interactions in Mn$_{12}$.
Recent experiments \cite{neutr,specheat,dmbar} 
show that 
the excitations with spin values ${\cal S}<10$ are 
rather close to the ground state: the distance is 
40--60 K (values differ in different reports). 
This is less than the energy of 
some states with the spin ${\cal S}=10$ 
(namely, the states ${\cal S}_z=0,\pm 1,\pm 2, \pm 3$), 
i.e.\ the lower states of the manifold ${\cal S}=9$ are 
lower than the 
higher states of the manifold ${\cal S}=10$. 
Thus, an adequate description of Mn$_{12}$ should 
account for the excitations with ${\cal S}<10$; i.e.\ 
the cluster should be considered as a many-spin system.

The total number of spin states in Mn$_{12}$ is large even for 
modern computers. But we can employ the 
fact that the exchange antiferromagnetic interactions $J_1$ 
(see Fig.\ \ref{figmn1}) are much larger than all 
the others \cite{gat1}, so corresponding pairs of ions Mn$^{3+}$ 
and Mn$^{4+}$ 
form dimers with the total spin $s=1/2$ (one of this pairs is 
designated in Fig.\ \ref{figmn1}, it includes ions $C$ and 
$D$). This model has already been 
successfully used for description of spin states of the 
cluster \cite{gat1,zvezdin}. Its validity is proven by 
megagauss-field experiments \cite{dotsenko}: 
the states of dimers with the spin $s$ higher than 1/2 
(excitations of dimers) come 
into play when the external magnetic field 
is about 400 T, i.e.\ the excitations of dimers 
have energy about 370 cm$^{-1}$. Analogously, 
the dependence of the magnetic susceptibility of the 
cluster versus temperature \cite{conf,gatteschi,gat1} shows 
that the dimer excitations contribute when 
temperature becomes as high as 150--200 K. 

Based on these data, we can analyze 
the domain of validity of the "dimerized" model. 
To do this, we note that the exchange interactions $J_2$, 
$J_3$ and $J_4$ mix 
the ground state of a dimer with the dimer excitations,
and the approximation of spin-1/2 dimers 
corresponds to the zeroth order perturbation theory with 
$1/J_1$ as an expansion parameter (similar approach has been used 
in Ref.\ \onlinecite{zvezdin}). 

To clarify this point, let 
us consider the level $a$ having, to the zeroth order, 
the energy $E_a$ with respect to the ground state.
Let us denote the distance between the ground state and the 
excitation of dimer as $E_{ex}\sim 370$ cm$^{-1}$.
The first-order correction 
to the energy of the level $a$ 
is of the order of $J'^2/(E_{ex}-E_a)$, where $J'$ is 
the magnitude of exchange interactions between dimers and 
nondimerized spins (see below).
Thus, accounting for the first-order corrections, the 
distance between the ground state and the level $a$ becomes 
$$
E_a'=E_a+C_a J'^2\left[ 1/(E_{ex}-E_a)-1/E_{ex})\right],
$$
where $C_a$ is a factor of order of unity, depending on 
the specific level $a$. 
As will be shown below, $J'$ is of order of 70 cm$^{-1}$;
so the first-order correction for the levels with 
energies about 70 cm$^{-1}$ is already considerable, 
of order of 4 cm$^{-1}$. This estimate, though being 
rough, gives the correct order of magnitude of the error 
introduced by the dimerized 8-spin model.

Moreover, this error restricts the region of temperatures 
where the dimerized model can be successfully applied. 
E.g., as our 
calculations show, to obtain the correct value of 
the magnetic susceptibility $\chi$ at the temperature $T$, we need 
to account for the levels with energies about $4\mbox{--} 5\,kT$. 
Obviously, the error in positions of these levels will 
introduce corresponding error in the dependence $\chi(T)$.
Its analytical evaluation is difficult, 
and the comparison of the results of calculations 
with the experimental data, performed in Sec.\ \ref{sectparam} 
is the better way to understand the temperature domain 
of validity of the dimerized model. 
As our results show, the dimerized model gives reasonable 
results for temperatures lower than about 50 K. 

Recalling that the temperatures below 30 K are of most 
interest, we conclude that the 
dimerized model is satisfactory for present needs of 
experimentalists.

\section{The spin Hamiltonian of M\lowercase{n}$_{12}$}
\label{sectham}

Thus, we consider the Mn$_{12}$ cluster as 
consisting of four "small" dimer spins $s=1/2$ and four 
"large" spins $S=2$ 
(corresponding to the four non-dimerized ions Mn$^{3+}$),
coupled by exchange interactions (see Fig.\ \ref{figmn2}). 
Moreover, we have to account
for the anisotropic relativistic interactions in the cluster, so
the Hamiltonian of the system can be written as:
\begin{equation}
\label{ham}
\nonumber
{\cal H}= -J\Bigl(\sum_i {\bf s}_i \Bigr)^2
  -J'\sum_{\langle k,l\rangle} {\bf s}_k {\bf S}_l + H_{\text{rel}},
\end{equation} 
where ${\bf s}_i$ are the spin operators of small dimer 
spins $s=1/2$, 
${\bf S}_l$ are spin operators of large spins $S=2$, and 
$H_{\text{rel}}$ denotes the part of the Hamiltonian describing 
relativistic interactions in the cluster. Summation in (\ref{ham})
is over pairs of spins coupled by exchange interactions. 
In the first term of the Hamiltonian we took into account 
that each small dimer spin is coupled with all the other 
small spins, so 
$2\sum{\bf s}_i{\bf s}_j=\left(\sum{\bf s}_i\right)^2$ up to 
an insufficient additive constant.

To zeroth order in $J_1$, the exchange integrals 
of the dimerized models 
are connected with the initial exchange parameters $J_2$, 
$J_3$ and $J_4$ as follows:
\begin{equation}
J=-J_2/2,\qquad J'=-J_3+2 J_4.
\end{equation}
Since the values of $J_2$, $J_3$ and $J_4$ are not known,
the parameters $J$ and $J'$ are to be determined from 
experimental data (see Sec.\ \ref{sectparam}). 

Furthermore, different types of relativistic anisotropic 
magnetic interactions possibly present in Mn$_{12}$ clusters
should be included in the Hamiltonian.
A large easy-axis anisotropy in the cluster is one of most
important features to be taken into account.
Generally, this anisotropy arises due to 
the single-site anisotropy of large spins 
(spins of Mn$^{3+}$ ions) and various kinds of 
anisotropic exchange.
We performed calculations for three basic types of 
easy-axis anisotropy in the cluster:
\begin{mathletters}
\begin{eqnarray}
\label{a}
H^1_{\text{rel}}&=& -K_z\sum_{i=1}^4 \left(S_i^z\right)^2,\\
\label{b}
H^2_{\text{rel}}&=& -J_{zz} \sum_{\langle i,j\rangle} s_i^z
 s_j^z,\\
\label{c}
H^3_{\text{rel}}&=& -J_{Zz} \sum_{\langle i,j\rangle} s_i^z
 S_j^z,
\end{eqnarray}
\end{mathletters}
where summations in (\ref{b}) and (\ref{c}) are over 
exchange-coupled pairs of spins. Anisotropy parameters 
($K_z$, $J_{zz}$ or $J_{Zz}$) have been 
chosen to give a correct value of
the zero-field splitting between the states ${\cal S}_z=\pm 10$ and
${\cal S}_z=\pm 9$ (14.4 K).
All three types of anisotropy give rather 
close energies of low-lying excitations (of energy less than
40 K), but higher excitations 
are reproduced best if the anisotropy is assumed to be of single-site
type (\ref{a}), so we can conclude that the easy-axis anisotropy
is primarily of single-site type. This result agrees with 
the conclusion drawn in Ref.\ \onlinecite{hfepr}. 
We will consider only this kind of anisotropy. 

Another potentially important sort of relativistic interaction 
is an in-plane anisotropy of large spins, i.e.\ $H_{\text{rel}}$ 
can include a contribution of the form:
\begin{equation}
\label{anitr}
H^4_{\text{rel}}= K_1 \left[\left(S_1^x\right)^2 + 
  \left(S_2^y\right)^2 +\left(S_3^x\right)^2 +
  \left(S_4^y\right)^2 \right],
\end{equation}
where the presence of fourth-order symmetry axis in the cluster 
is directly taken 
into account. The small spins $s=1/2$ are excluded since 
$\left(\sigma_x\right)^2=\left(\sigma_y\right)^2=
\left(\sigma_z\right)^2=1$ for Pauli 
matrices $\sigma_x$, $\sigma_y$ and $\sigma_z$; and only 
spins of non-dimerized Mn$^{3+}$ ions 
give a nontrivial contribution. These ions are surrounded by
eight oxygen ions forming a distorted octahedron. The
axes of oxygen octahedra 
are significantly tilted from the $c$-axis of the cluster, 
therefore, this term can be relatively large, even 
comparable to the easy-axis anisotropy. But, surprisingly, 
our results show that this kind of interaction gives negligible 
effect, except for trivial renormalization of the easy-axis anisotropy
constant $K_z$ in (\ref{a}). If we account for 
this renormalization, 
the positions and the wave functions of excited levels
remain almost unaffected even for $K_1=3 K_z$ (i.e., for the 
in-plane anisotropy three times larger than the easy-axis one).
Thus, this kind of interaction can be 
excluded from further considerations.

Another important interaction 
is Dzyaloshinsky-Morya (DM) antisymmetric exchange. 
To our knowledge, the possible 
presence of DM-interactions in Mn$_{12}$ was first suggested in
Ref.\ \onlinecite{dmbar}, but little attention has been 
paid until now. Our results show that these interactions are, indeed,
very important and have rather large magnitude. 

A pair of ions coupled by DM-interaction is described by 
the Hamiltonian
\begin{equation}
H_{\text{DM}}= {\bf D}\cdot [{\bf S}_1\times {\bf S}_2],
\end{equation}
and the magnitude of the DM-vector $\bf D$ can be estimated as
\cite{yosida} $D\sim \lambda A$, 
where $A$ is the isotropic (nonrelativistic) exchange 
coupling between ions and
$\lambda$ is the spin-orbit coupling constant (which is rather
small for transition ions). For comparison,
the magnitude of easy-axis anisotropy is estimated as
$K_z\sim \lambda^2 A$, 
i.e.\ is of next order of smallness in comparison with $D$.
Thus, the DM-interactions in Mn$_{12}$
can be expected to be important.

In the 8-spin model of the cluster there are DM-interactions 
of two kinds:
\begin{mathletters}
\begin{eqnarray}
\label{dma}
H_{\text{DM}}&=&\sum_{\langle i,j\rangle} {\bf D}^{i,j} \cdot
  [{\bf s}_i\times {\bf S}_j]\\
\label{dmb}
H^1_{\text{DM}}&=&\sum_i\sum_j {\bf D}_1^{i,j} \cdot
  [{\bf s}_i\times {\bf s}_j].
\end{eqnarray}
\end{mathletters}
Summation in (\ref{dma}) is over exchange-coupled 
pairs of spins; summation in (\ref{dmb}) is over all 
pairs of dimer spins, since all dimer spins interact 
with each other. 
We studied both kinds of DM-interaction and found that the
second kind, i.e. $H^1_{\text{DM}}$ involving small spins 
can be neglected. Therefore,
we can neglect the interactions of the type (\ref{dmb}).

The crystal field in Mn$_{12}$, governing the 
DM-interactions, possesses certain symmetry elements, thus 
imposing restrictions on the values of ${\bf D}^{i,j}$. 
It is reasonable (and rather standard \cite{yosida}) to 
assume that the crystal field is determined mainly 
by the oxygen octahedra surrounding manganese ions in 
the cluster; so the symmetry of the crystal field is 
governed by the mutual arrangement of the oxygen octahedra. 
The following two symmetry elements are of interest for us. 
The first one is the fourth-order rotary-reflection 
axis \cite{gatteschi,gat1} 
parallel to the $c$-axis of the cluster. This symmetry is
obviously preserved in the 8-spin model of the cluster,
so the two DM-vectors ${\bf D}^{1,5}$ and ${\bf D}^{1,8}$ (see 
Fig.\ \ref{figmn2}) define all the other ${\bf D}^{i,j}$.
The other element of symmetry is the mirror plane $\rho$ 
parallel to the $z$-axis passing through the ions $C$
and $D$ (see Fig.\ \ref{figmn1}).
The oxygen octahedra surrounding the ions $A$
and $B$ (see Fig.\ \ref{figmn1}) are invariant with a good
degree of accuracy \cite{mynote} 
with respect to reflection in the plane $\rho$
(inspection of the structure data supplied in Refs.\
\onlinecite{gatteschi,gat1} 
shows this); this symmetry is also preserved in the 8-spin
model. 
Thus, the vector ${\bf D}^{1,8}$ (Fig.\ \ref{figmn2})
defines
all the other DM-vectors in the Hamiltonian (\ref{dma}):
\begin{mathletters}
\label{dmvec}
\begin{eqnarray}
\label{dmvecx}
D_x^{1,8}&=&-D_x^{1,5}=D_y^{2,5}=-D_y^{2,6}=-D_x^{3,6}=D_x^{3,7}\\
  \nonumber
  &&=-D_y^{4,7}=D_y^{4,8},\\
\label{dmvecy}
D_y^{1,8}&=&D_y^{1,5}=-D_x^{2,5}=-D_x^{2,6}=-D_y^{3,6}=-D_y^{3,7}\\
  \nonumber
  &&=D_x^{4,7}=D_x^{4,8},\\
\label{dmvecz}
D_z^{1,8}&=&-D_z^{1,5}=D_z^{2,5}=-D_z^{2,6}=D_z^{3,6}=-D_z^{3,7}\\
  \nonumber
  &&=D_z^{4,7}=-D_z^{4,8}.
\end{eqnarray}
\end{mathletters}
Obviously, any other vector ${\bf D}^{i,j}$ can be taken as a 
basis instead of ${\bf D}^{1,8}$.
No other symmetry elements allow for further reduction,
so DM-interactions in Mn$_{12}$ are described using three 
parameters: $D_x^{1,8}$, $D_y^{1,8}$ and $D_z^{1,8}$. Below,
these parameters are denoted simply as $D_x$, $D_y$ and $D_z$.

As our results show, in the DM-Hamiltonian (\ref{dma}) the terms
proportional to $D_y$ produces negligible matrix elements 
(a few percent in comparison with other terms). It occurs 
due to symmetry reasons: the inspection
of the relations (\ref{dmvec}) shows that the components
$D_x$ and $D_z$ transform antisymmetrically with respect to 
reflection in the plane 
$\rho$, but the component $D_y$ 
transforms symmetrically. 
The matrix elements of the terms proportional to $D_y$ 
nearly cancel each other, leading to 
negligible matrix elements. Therefore,
these terms are excluded from consideration and we 
set $D_y=0$ with negligible error. 

Finally, having studied all the interactions described above, we can 
write down the Hamiltonian of the cluster in the following form:
\begin{eqnarray}
\label{hamilton}
{\cal H}&=& -J\Bigl(\sum_i {\bf s}_i \Bigr)^2
  -J'\sum_{\langle k,l\rangle} {\bf s}_k {\bf S}_l
  -K_z\sum_{i=1}^4 \left(S_i^z\right)^2\\ \nonumber
  && + \sum_{\langle i,j\rangle} {\bf D}^{i,j} \cdot
     [{\bf s}_i\times {\bf S}_j],
\end{eqnarray}
where the DM-vectors ${\bf D}^{i,j}$ obey the relations (\ref{dmvec})
with the parameter $D_y^{1,8}\equiv D_y=0$.

\section{Review of relevant experimental results}
\label{sectexpres}

At present, 
data of various experiments on magnetic molecules Mn$_{12}$Ac 
are available, including 
the temperature dependence of the effective magnetic moment of
the cluster $\mu_{\text{eff}}(T)$ \cite{gatteschi,gat1,neutr}, 
the 
results of EPR experiments \cite{hfepr,boom}, dynamic 
susceptibility measurements \cite{her}, 
inelastic 
neutron scattering data \cite{neutr} and specific heat data 
\cite{specheat}. Unfortunately, only few of these data
can be used for determining the parameters of the 8-spin
Hamiltonian for Mn$_{12}$ clusters. 

Recent high-frequency EPR experiments \cite{hfepr} 
refined the description of
the easy-axis anisotropy of the cluster and showed that 
the anisotropy Hamiltonian in the single-spin model 
can be approximated as follows:
\begin{eqnarray}
\label{hfappr}
H&=&\alpha{\cal S}_z^2 + \beta{\cal S}_z^4 +
  \gamma ({\cal S}_+^4+{\cal S}_-^4),\\
  \nonumber
  \alpha&=&-0.56\text{ K},\quad\beta=-11.08\cdot 10^{-4}\text{ K},\\ 
  \nonumber
  \gamma&=&2.88\cdot 10^{-5}\text{ K},
\end{eqnarray}
where
${\cal S}_z$, ${\cal S}_+$ and ${\cal S}_-$ denote 
the operators of the total spin of the cluster.
It means, in terms of a many-spin approach, that 
the energies of the low-lying levels with spin ${\cal S}=10$ 
obey Eq.\ (\ref{hfappr}).
It is worth noting that the derived values of quartic 
corrections $\beta$ and $\gamma$ are rather large and,
as our calculations show (see below), seem
to be poorly explained using the single-spin description of
Mn$_{12}$, i.e.\ when accounting only for the states belonging
to the ${\cal S}=10$ manifold. Our results show that the 
excited levels with ${\cal S}<10$ are necessary to give reasonable 
values for the quartic corrections. 

Another set of results, very useful for elucidating the 
many-spin interactions in Mn$_{12}$Ac, is the neutron scattering
results supplied in Ref.\ \onlinecite{neutr}. 
The experiments have been performed at very low temperatures
(mostly, 1.5 K to 2.5 K), where only the lowest levels 
${\cal S}_z=\pm 10$ 
are populated. Since the selection rule for neutron 
scattering is $\delta{\cal S}_z=0,\pm 1$, only the levels 
with ${\cal S}_z=\pm 9$ can give rise to scattering peaks 
(the levels with ${\cal S}_z=\pm 11$ have too large energies 
\cite{dotsenko} and can be excluded).

Results of these experiments can be summarized 
as follows. A prominent peak of spin origin 
at about 0.3 THz has been 
detected and attributed to the transitions to the levels
with ${\cal S}=10, {\cal S}_z=\pm 9$, in excellent agreement with 
all previous data (0.3 THz corresponds to about 14.4 K).
At higher energies, two sets of peaks have been detected 
around 1.2 THz and 2.0 THz. The fitting proposed 
in Ref.\ \onlinecite{neutr} gives two peaks in the first set
(at energies about 57 K and 66 K) and three peaks in the
second set (at energies 90 K, 96 K and 105 K); but authors 
indicate clearly that possibly more peaks are present 
(most likely, three peaks in the first set and four or five in
the second). Another interesting detail of 
the neutron scattering spectra is a very broad mode 
situated at about 0.2 THz; this mode disappears when 
the temperature is less than about 2 K.

The authors have not managed to interpret these features, 
except for the peak at 0.3 THz. They pointed out that there is 
particular difficulty in interpretation of the peaks at 1.2--1.3 THz:
the model they used for susceptibility fitting gives 
two degenerate levels ${\cal S}=9$ at about 33 K, an obvious
contradiction with the neutron scattering spectrum. 
We show below that the 8-spin model developed here can
overcome these difficulties and gives correct positions 
for neutrons peaks at 1.2 THz along with a correct 
description of the susceptibility data.

Thus, we found the following experimental 
results to be relevant for the purpose of a quantitative 
description of the Mn$_{12}$ clusters.
The distance between the ground state and the first 
excited level(s) is 14.4 K. The energies (the anisotropy 
splittings) of 
the low-lying levels, belonging to the ${\cal S}=10$ manifold,
obey formula (\ref{hfappr}).
There are two or three neutron 
peaks around 60--70 K, two of them are situated at 57 K and
66 K. Also, there are up to five peaks around 100 K, three 
of them are at 90 K, 96 K and 105 K. 
The temperature dependence of the susceptibility (or, equivalently,
the dependence $\mu_{\text{eff}}(T)$) has the form 
displayed in Refs.\ \onlinecite{gatteschi,gat1,neutr} and 
Fig.\ \ref{figsusc}.

The other experimental results, though providing 
important information about Mn$_{12}$, are much less 
suitable for our purposes (to a large extent, because 
different, {\it a priori\/} equally probable, 
interpretations are possible).

\section{Numerical calculations and parameters of
the 8-spin model. Comparison with experiment}
\label{sectparam}

Having derived the spin Hamiltonian for the 8-spin model of 
Mn$_{12}$, we attempted to extract its parameters from the 
relevant experimental data. 

We used the following two-step 
numerical scheme. At the first step, the relativistic 
term $H_{\text{rel}}$ has been neglected resulting in 
an isotropic exchange Hamiltonian. The eigenstates 
of this Hamiltonian are degenerate with 
respect to ${\cal S}_z$. Thus, it is sufficient to take 
into account only the states with ${\cal S}_z=0$,
so the exchange Hamiltonian (represented by 
a matrix 1286$\times$1286) has been diagonalized 
within the subspace spanned by these states.
Then, at the second step, 
the relativistic anisotropic interactions have been
taken into account. Among the states obtained at 
the first step (having ${\cal S}_z=0$), 
we retain only those with the energy 
less than $E_{\text{cut}}$ (a sufficiently large 
value for this parameter has been chosen) and generate 
the corresponding states with different ${\cal S}_z$ 
(basis states). Then 
the complete Hamiltonian (\ref{hamilton}) has been 
diagonalized within the subspace spanned by the 
generated basis states.
Calculations with different values of $E_{\text{cut}}$ 
have been performed to assure that the 
positions of lower levels are obtained with 
desired accuracy. Typical values of $E_{\text{cut}}$ 
were about 250 K: the levels with higher energies 
are not worth including due to the limited 
accuracy of the 8-model itself (see Sec.\ \ref{sectmod}). 

Based on the procedure described above, the fitting 
of relevant experimental data (Sec.\ \ref{sectexpres})
has been made and 
the possible parameters of the 8-spin Hamiltonian
determined. 

Neutron scattering data are of primary interest for us. 
We focus our attention on the positions of the neutron 
peaks, since the amplitudes depend strongly on 
details of the experiments. 
We first assume an ideal experiment,
where the resolution of 
the setup is infinite and the neutrons with 
all possible scattering vectors are detected (i.e.,
the detector has infinite 
aperture). In this case, at zero temperature, 
the cross-section of 
neutron scattering at the energy $E$ is \cite{neutr,white}
\begin{eqnarray}
\label{cross}
\sigma (E)&=& \int_{\bf R^3} d^3{\bf q}\;A F^2(q) 
  \sum_{a,b}\left(\delta_{a,b} - 
  q_a q_b/q^2\right)\\
 \nonumber
 &&{}\times \sum_{m,n} \exp{[i{\bf q (r_{\it m}-
  r_{\it n})}]}\\
 \nonumber
 &&{}\times \sum_{\psi} 
  \langle 0| S^a_m |\psi\rangle 
  \langle \psi| S^b_n |0\rangle\; 
  \delta (E(\psi)-E),
\end{eqnarray} 
where $A$ is a constant, $F(q)$ is the form-factor of 
manganese ions, $\bf q$ is a scattering vector, $n$, $m$ 
enumerate different ions, and $a$, $b$ refer to 
the Cartesian coordinates ($x$, $y$ and $z$). The integration 
is performed over all vectors $\bf q$. $E(\psi)$ denotes 
the energy of the state $\psi$, $|0\rangle$ denotes the 
ground state, which is the only one populated at zero 
temperature.
The transitions with $\Delta{\cal S}_z=0,+1$ can be neglected 
since there is only one state ${\cal S}_z=10$ and no states 
${\cal S}_z=11$. In this case, the total cross-section 
(\ref{cross}) is proportional to the quantity 
\begin{equation}
\label{vis}
V=\sum_i \left| \bigl\langle \phi^{(9)}_i 
  \bigl|\psi\bigr\rangle \right|^2,
\end{equation}
where the state $\psi$ has the energy $E(\psi)=E$ with 
respect to the groundstate; i.e., the state $\psi$ is 
the final state of the neutron scattering process 
and gives rise to a neutron peak at the energy $E(\psi)$ of 
the amplitude proportional to $V$. 
The summation in (\ref{vis}) is performed over 
all basis levels 
having ${\cal S}_z=9$; these levels are denoted 
as $\phi^{(9)}_i$. Equation (\ref{vis})
expresses the simple fact that only the transitions with 
$\Delta{\cal S}_z=-1$ are allowed in the neutron 
scattering process, since the transitions with 
$\Delta{\cal S}_z=0,+1$ are absent. 
Below, the quantity $V$ is referred to 
as a normalized cross-section for the level $\psi$.
Our results show that $V$ discriminates 
easily the eigenstates which can give rise to noticeable 
neutron scattering peaks.

Furthermore, the values of parameters $\alpha$ and $\beta$
describing the easy-axis 
anisotropy in Eq.\ (\ref{hfappr}), 
have been taken into account in determination of the 
cluster parameters. 
The energies of the five lowest 
levels, having spin ${\cal S}=10$, have been approximated 
by a fourth-order polynomial, following Eq.\ (\ref{hfappr}),
and the coefficients $\alpha$ and $\beta$ have 
been extracted and compared to the experimental data. 

As a result of calculations, the following three sets of 
the cluster parameters have been found to provide 
the best fitting of experimental data: 
\begin{description}
\item[\rm Set A:]
$J=0$, $J'=105$ K, $K_z=5.69$ K, $D_z=-1.2$ K, $D_x=25$ K;
\item[\rm Set B:]
$J=23.8$ K, $J'=79.2$ K, $K_z=5.72$ K, $D_z=10$ K, $D_x=22$ K;
\item[\rm Set C:]
$J=41.4$ K, $J'=69$ K, $K_z=5.75$ K, $D_z=10$ K, $D_x=20$ K.
\end{description}
The positions of neutron peaks calculated for these sets of parameters 
are presented in Fig.\ \ref{figneut}. 
The graphs show the dependence of normalized cross-section
vs. the level energy. It is seen from these figures, 
that the normalized cross-section is extremely small 
(less than $10^{-2}$) for most of levels, and 
only few states can give rise to 
noticeable neutron scattering peaks.
Moreover, to facilitate the analysis of the data for reader, the 
positions of neutron peaks are listed in the Table 
\ref{tabneut}. The values of the easy-axis anisotropy 
parameters $\alpha$ and $\beta$ are listed in the 
Table \ref{tabanis}. 

As the results show, each of the parameter sets reproduces 
reasonably well its own portion of the experimental results. 
All the sets give reasonably good positions of 
the low-energy neutron 
peaks at 0.3 THz (14.4 K), 1.19 THz (57 K) and 1.38 THz (66 K). 
The parameter 
set A also gives the values of anisotropy parameters 
$\alpha$ and $\beta$, rather 
close to the experimental ones, 
but the neutron peaks corresponding to higher energies (around 
2 THz) are reproduced poorly. 
The parameter sets B and C give correctly only the 
order of magnitude of $\alpha$ and 
$\beta$, but reproduce better the positions of the 
high-energy neutron peaks. 

Finally, the temperature dependence of the effective magnetic 
moment $\mu_{\text{eff}}$ of the cluster 
has been calculated for all three
sets of parameters using the formula
\begin{equation}
\label{mueff}
\mu_{\text{eff}}\,(T)\equiv \sqrt{3\chi(T)\cdot k T},
\end{equation}
where $\chi$ is the susceptibility of the cluster, 
$k$ is Boltzmann's constant and $T$ is the temperature. 
The susceptibility $\chi(T)$ has been calculated 
in a way reproducing the experimental procedure. 
The Zeeman term, describing the effect of an external field 
has been introduced into the Hamiltonian (\ref{hamilton}). 
The field magnitude $H=1$ mT has been chosen following 
Ref.\ \onlinecite{neutr}. The resulting Hamiltonian has 
been diagonalized, and the component of the cluster 
spin ${\cal S}_{\text{H}}$ along the
field has been calculated by means of quantum-statistical 
averaging over the Gibbs canonical ensemble. This routine 
has been repeated several times for different orientations 
of the field, and the obtained values of ${\cal S}_{\text{H}}$ 
have been averaged. 
It corresponds to a powder sample measurements, 
when the crystallytes are randomly oriented with respect to 
the field. 
The susceptibility $\chi$, following 
a standard experimental procedure, has been calculated as a 
ratio of the resulting average cluster spin to the 
field magnitude $H$. Finally, the value $\mu_{\text{eff}}$ 
has been obtained applying Eq.\ (\ref{mueff}).

The curves $\mu_{\text{eff}}$ calculated for the three
sets of cluster parameters, are presented in Fig.\ 
\ref{figsusc} along with the experimental data. 
All the sets give almost coinciding curves, 
and below 50 K the agreement with experiment is good. 
The region of temperatures higher than 50 K 
can not be reproduced satisfactorily:
as our test calculations showed, 
to obtain the correct value of 
effective moment $\mu_{\text{eff}}$ at the temperature $T$, we need 
to account for the levels with energies about $4\mbox{--} 5\,kT$. 
When calculating the curves presented, 
only the levels with energies 
less than 250 K have been taken into account, 
thus restricting the correctly described temperature region.

\section{Qualitative analysis of the results and 
interpretation of experimental data}
\label{sectinter}

At present, having rather limited number of the relevant 
experimental data, it is hard to distinguish between the 
parameter sets A, B and C. The easy-axis anisotropy 
parameters $\alpha$ and $\beta$ are obtained with good 
precision in EPR experiments, but the magnetization 
measurement data \cite{dmbar} suggest other values for 
these parameters; so, comparison of the experimental 
values of $\alpha$ and $\beta$ with our results 
can not serve as a definitive basis for judgement. 
Also, the quality of the description of 
the high-energy neutron peaks can not be decisive, since 
the disagreement 
can be attributed to the limited accuracy of the dimerized 
8-spin model itself. 
Megagauss-field experiments \cite{dotsenko}, along with 
careful measurements of the low-energy 
peaks (around 1.2 THz) and fitting of their amplitudes seems 
to be a promising strategy for future investigations.

Nevertheless, the results already obtained provide 
new important information about the role of 
many-spin interactions in Mn$_{12}$ clusters. In 
this section we focus our attention on the qualitative 
consideration of the results obtained and 
discuss the interpretation of the experimental 
data. 

First, it is worthwhile to note that the consideration of 
states with spin ${\cal S}$ less than 10 
leads to rather large quartic corrections to the 
energy of easy-axis anisotropy. If these excited states 
are not taken into account, i.e.\ if only the states 
with ${\cal S}=10$ are included, the value of $\beta$ 
is of order of 10$^{-5}$ K. 

Another important fact is the large magnitude of the 
Dzyaloshinski-Morya (DM) interactions in the cluster Mn$_{12}$. 
In our opinion, this can be attributed to the low symmetry 
of the cluster. Indeed, the strength of the DM-interaction 
is governed to a large extent by asymmetry of crystal field acting on
the interacting ions \cite{yosida}. An instructive 
example is provided in
Ref.\ \onlinecite{lacr}: the DM-interaction can 
emerge for ions located at the 
surface of a magnet, even though these interactions are prohibited for
ions in the bulk of the magnet.
In some sence the Mn$_{12}$ molecule possesses "surface" everywhere,
and the symmetry of the crystal field is rather low.

The presence of the large Dzyaloshinsky-Morya term in the 
Hamiltonian provides a key to an explanation of the 
neutron scattering data.
First, the DM-terms lead to the appearance of the 
two neutron peaks around 1.2 THz. If these 
terms are absent, there are two degenerate levels with 
${\cal S}_z=9$ around 1.2 THz. Among all the 
interactions we considered (see Sec.\ \ref{sectham}), only
the DM-interaction can lift this 
degeneracy and provide a large splitting (about 9 K), 
as observed in 
experiments. Similarly, according to our calculations, 
several peaks around 2 THz appear only due to DM-interactions. 

The origin of the peak at 0.3 THz has been completely 
explained in Ref.\ \onlinecite{neutr}: it appears 
because of easy-axis anisotropy splitting the levels 
with different ${\cal S}_z$. Our results agree with 
this conclusion. 

An interesting feature in the neutron scattering data 
is the broad mode situated at 0.2 THz.
No states of this energy have been observed, e.g., in 
EPR-experiments \cite{gat1,hfepr,boom}. Our 
calculations also show no states with the energy 0.2 THz
(or, equivalently, about 10 K). 
In our opinion, this mode is caused by an 
interaction of Mn$_{12}$ clusters with the 
dissipative environment. Due to this interaction, 
each level broadens (nonuniform broadening), 
forming a quasiband of finite width $\delta E$ 
(see Refs.\ \onlinecite{hartman,my} for details).
The value of $\delta E\approx 2$ K can be 
estimated from the single-crystal 
hysteresis measurements \cite{jumps1}. 
Transitions between the two quasibands take place,
so, along with the peak at 0.3 THz (14.4 K), 
a broad mode of inter-quasiband transitions appears. 
It happens when different states in the quasibands 
are populated, i.e.\ 
at the temperatures of order of $\delta E\approx 2$ K,
which agrees with experiment. 
Energies of the inter-quasiband transitions are reduced 
by the value about $2\cdot\delta E\approx 4$ K, so 
the corresponding neutron scattering mode is situated around 
$E_b=14.4 \text{ K} - 2\cdot\delta E\approx 10.4$ K, 
or, equivalently,
around 0.2 THz, in agreement with experiment. At increasing 
temperatures, the occupancy of quasibands becomes 
more uniform, so the intensity of the broad mode increases along 
with the decrease in intensity of the peak at 14.4 K. 
This behavior also agrees with experiment.
However, this qualitative explanation can not 
be considered as sufficient, 
and a rigorous quantitative treatment is necessary. 
Such a treatment constitutes a separate 
physical problem to be investigated in the future.

\section{Dzyaloshinsky-Morya interactions and the 
relaxation of magnetization}
\label{sectrelax}

Strong Dzyaloshinsky-Morya interactions can play an 
important role in relaxation of magnetization in Mn$_{12}$ 
clusters. 
The possible importance of 
DM-interactions for the relaxational properties 
of Mn$_{12}$ was already mentioned in Ref.\ \onlinecite{dmbar}, 
but no analysis was made. 
Below, the relation between the DM-interaction and 
relaxation is analyzed qualitatively. 

The relaxation time in Mn$_{12}$ is very long, 
about 2 months at 2 K. 
At relatively high temperatures the magnetization relaxation 
time $\tau$ follows the Arrhenius law 
$\tau=\tau_0\exp{(-\Delta E/kT)}$
with $\tau_0=2.4\cdot 10^{-7}$ sec and the effective barrier 
$\Delta E=61$ K \cite{gat1}. At low temperatures the
situation changes: the 
experimental results presented in Refs.\ \onlinecite{barrel,park}
show that $\tau$ saturates below the crossover temperature 
$T_c=2$ K, 
which is attributed to the onset of "ground state - to - ground state" 
tunneling. Such a high value of the crossover temperature 
poses certain 
problems, since at present the nature of magnetic relaxation 
in Mn$_{12}$ is unknown.
Important information has been supplied by hysteresis 
measurements \cite{jumps,jumps1}: it has been directly 
shown that the relaxational mechanism in Mn$_{12}$ should 
allow for transitions between the states $|{\cal S}_z\rangle$ 
and $|{\cal S}_z\pm 1\rangle$. 

The relaxation of magnetization has been extensively studied 
\cite{garanin,politi,stamp} within the single-spin model 
of Mn$_{12}$. The Hamiltonian (\ref{hfappr}) of this model 
provides transitions with $\Delta{\cal S}_z=\pm 4$ only, 
and an external transversal field $H_x$ is necessary 
to allow the $\Delta{\cal S}_z=\pm 1$ transitions. 
Different sources of this field have been considered: 
a hyperfine field induced by nuclear spins \cite{garanin} 
and a dipole-dipole field induced by other clusters \cite{stamp}. 
There are certain difficulties facing these interpretations. 
E.g., as reported in Refs.\ \onlinecite{hartman,sessoli}, the 
relaxation rate increases on dilution of Mn$_{12}$ 
in solution, 
although the dipole-dipole interactions between the 
clusters decreases; this seems to be difficult to explain 
with the relaxation mechanism based on intercluster 
dipole-dipole interactions \cite{stamp}.
The other mechanism, based on the hyperfine fields \cite{garanin}
can not easily explain the high value of the crossover temperature. 

Strong Dzyaloshinsky-Morya interactions constitute 
another source of magnetic relaxation in Mn$_{12}$. 
It obviously allows the $\Delta{\cal S}_z=\pm 1$ transitions. 
On solution, the dipole-dipole fluctuating fields 
between the clusters decrease, thus decreasing the 
decohering influence of the environment, so the relaxation 
rate increases. This agrees with experimental 
results \cite{hartman}. Relatively large magnitude 
of DM-interactions can, in principle, explain the 
high value (2 K) of the crossover temperature.
This shows that the quantitative study of DM-based 
relaxation in Mn$_{12}$ is important. 

Nevertheless, we emphasize that, to our knowledge, 
current information is 
not sufficient for judgement in favor of some single 
relaxation mechanism. At present, all of them can 
be considered as equally probable, and the possibility 
of a combination of different mechanism exists. 

Independent of the problem of relaxation in 
Mn$_{12}$, DM-interactions present an interesting and 
important mechanism for magnetic relaxation. It can 
be significant, for example, in nanosized particles, 
where large DM-interactions can arise 
due to reduced symmetry at the surface \cite{lacr}. 
The relaxation based on this interaction is 
rather unusual: it is related not to the potential 
barrier created by easy-axis anisotropy, but to 
the barrier created by isotropic exchange. The point 
is that the Dzyaloshinsky-Morya term $H_{\text{DM}}$ couples 
the states with different values of total spin ${\cal S}$, 
i.e.\ 
\begin{equation}
\langle {\cal S}, {\cal S}_z| H_{\text{DM}} 
 |{\cal S}\pm 1, {\cal S}_z\pm 1\rangle\neq 0. 
\end{equation}
This feature differs drastically from standard consideration 
of relaxation in small particles, where only the 
anisotropy barrier is usually taken into consideration. 

Isotropic exchange interactions are usually rather strong, 
but it does not necessarily means that the DM-based relaxation 
is negligible. E.g., in spin-frustrated systems the height 
of the exchange barrier can be considerably reduced. Moreover, 
it can be significant in certain non-frustrated systems. 
For the relaxation based on DM-interactions, the relaxation 
time should be governed by the ratio $D/A$, where $D$ is the 
absolute value of Dzyaloshinsky-Morya vector and $A$ is the 
exchange integral. On the other hand, for "conventional" 
relaxation mechanism, when the anisotropy barrier is 
overcome, the ratio $U/K$ governs the 
relaxation rate, where $K$ is the anisotropy constant and 
$U$ is the strength of the interaction between the spins 
and their dissipative environment (e.g., spin-phonon 
coupling constant). Situations with $D/A \gg U/K$ are 
not impossible. 

Therefore, quantitative investigations of 
relaxation mechanisms based on DM-interactions is of 
great interest and importance.

\section{Summary}
\label{sectsum}

In the present work, we 
have performed an extensive study of spin excitations in Mn$_{12}$,
explicitly accounting for its many-spin internal structure. 
The dimerized 8-spin model of the Mn$_{12}$ clusters 
\cite{gat1} has been used. 
Along with isotropic exchange coupling, various kinds of
anisotropic relativistic interactions have been studied:
anisotropic exchange coupling between the cluster ions, 
single-site anisotropies of easy-axis and in-plane type, and 
various kinds of Dzyaloshinsky-Morya (DM) interactions.
Surprisingly, most of these interactions play only a minor role.

As a result, we propose a basic many-spin Hamiltonian 
which includes isotropic exchange couplings, 
single-site anisotropies of easy-axis type and 
DM-interactions between the cluster spins. 
Three possible sets of parameters are determined 
from the relevant experimental data. 
The results of our calculations 
reproduce satisfactorily various experimental results, such as 
positions of neutron scattering peaks, high-frequency EPR 
data and the experimental dependence of the magnetic susceptibility 
on temperature. 

In particular, our results suggest rather strong 
Dzyaloshinsky-Morya interactions are present 
in the Mn$_{12}$ cluster. We have discussed qualitatively 
the possible relation of these interactions to the 
unusual magnetic relaxational 
properties of Mn$_{12}$. We emphasized that 
DM-interactions present interesting 
relaxation mechanisms worth further investigations.

\section*{Acknowledgments}
Authors would like to thank A.\ K.\ Zvezdin, B.\ Barbara, D.\ Garanin 
for many helpful discussions. 
This work was partially carried out at the Ames Laboratory, which 
is operated for the U.\ S.\ Department of Energy by Iowa State 
University under Contract No.\ W-7405-82 and was supported by 
the Director for Energy Research, Office of Basic Energy Sciences 
of the U.\ S.\ Department of Energy. This work was partially supported 
by Russian Foundation for Basic Research, grant 98-02-16219.

\begin{figure}
\caption{Schematic plot of the Mn$_{12}$ cluster. Small black 
circles represent Mn$^{4+}$ ions, large white circles --- Mn$^{3+}$
ions. Different types of lines connecting the ions (solid, dashed,
dotted and dash-dotted) correspond to different types of exchange 
interactions ($J_1$, $J_2$, $J_3$ and $J_4$).}
\label{figmn1}
\end{figure}

\begin{figure}
\caption{A schematic plot of the 8-spin system representing
the Mn$_{12}$ cluster. White large circles represent large spins 
($S=2$), and dark small squares represent small dimer spins 
($s=1/2$).}
\label{figmn2}
\end{figure}

\begin{figure}
\caption{Dependence of the normalized cross-section 
vs. level energy (in K), calculated for the three sets 
of the cluster parameters (A, B and C, see text).
The levels producing noticeable neutron peaks can be 
easily discriminated from the others.}
\label{figneut}
\end{figure}

\begin{figure}
\caption{Temperature dependence of the effective magnetic 
moment of the cluster $\mu_{\text{eff}}$ (in Bohr's magnetons). 
Results of 
calculations with the three sets of parameters are 
shown: the set A (solid line), the set B 
(dashed line) and the set C (dotted line). 
Large solid squares represent experimental data.
The results of calculations with the sets A and B are 
very close to each other, and the corresponding curves merge
on the figure.}
\label{figsusc}
\end{figure}

\begin{table}
\caption{The positions of neutron peaks: 
comparison between experimental data and calculated results. 
Calculations have been made for the three possible 
sets of the cluster parameters (A, B and C, see text).
The levels with normalized cross-section more than 0.05 and 
energy less than 130 K are included in the table. }
\begin{tabular}{ldddd}
     &Experiment & Set A & Set B & Set C \\
\tableline
                &14.4 K & 14.4 K & 14.4 K & 14.4 K\\
                &       &        &        & \\
Low-energy      & 57 K & 58.2 K & 55.2 K & 56.7 K\\
peaks (1.2 THz) & 66 K & 66.0 K & 66.7 K & 67.0 K\\
                & maybe, &      & 67.4 K & 67.9 K\\
                & more & 76.6 K &        & 75.7 K\\
                &      &        &        &       \\
High-energy     & 90 K & 124.4 K & 98.3 K & 88.8 K\\
peaks (2 THz)   & 96 K & 124.9 K & 105.1 K & 110.9 K\\
                & 105 K & 126.4 K & 110.4 K & 122.1 K\\
                & maybe, & 127.1 K & 122.1 K & \\
                & more   &         &         & \\
\end{tabular}
\label{tabneut}
\end{table}

\begin{table}
\caption{Parameters $\alpha$ and $\beta$ of the 
easy-axis anisotropy: comparison 
between experimental data and calculated results.
Calculations have been performed for the 
three possible sets of the cluster parameters (A, B and C, 
see text).}
\begin{tabular}{ldddd}
     &Experiment & Set A & Set B & Set C \\
\tableline
$\alpha$ (K) & $-$0.56 & $-$0.63 & $-$0.68 & $-$0.67 \\
$\beta$ (mK) & $-$1.11 & $-$0.7 & $-$0.45 & $-$0.49 \\
\end{tabular}
\label{tabanis}
\end{table}


\begin{references}
\bibitem{mynote} Although this symmetry is not exact, 
  the deviation from the symmetry is small. Such approximate 
  symmetries are widely used in physics: e.g., the crystal 
  lattice of ferromagnetic iron is often considered as cubic, though 
  actually it is tetragonally distorted.
\end{references}
\end{document}